\documentclass[]{aa}  

\usepackage{graphicx}

\usepackage{txfonts}
\usepackage{hyperref}

\usepackage{amsmath}
\usepackage{comment}
\usepackage{cleveref}
\usepackage{subfigure}
\usepackage{xcolor}
\usepackage{placeins}
\usepackage{capt-of}

\usepackage{savesym}
\savesymbol{tablenum}
\usepackage{siunitx}
\restoresymbol{SIX}{tablenum}

\newcommand{\revision}[1]{{#1}}

\begin{document}

    \title{ Can gap-edge illumination excite spirals in protoplanetary disks?}

   \subtitle{Three-temperature radiation hydrodynamics and NIR image modeling}

   \author{Dhruv Muley \inst{1}
        \and Julio David Melon Fuksman \inst{1}
        \and Hubert Klahr \inst{1}          }

   \institute{Max-Planck-Institut f\"ur Astronomie, Königstuhl 17, Heidelberg, DE 69117\\
              \email{muley@mpia.de}            }

   \date{Received 18 July 2024; accepted 28 July 2024}

% \abstract{}{}{}{}{} 
% 5 {} token are mandatory
 
  \abstract{The advent of high-resolution, near-infrared instruments such as VLT/SPHERE and Gemini/GPI has helped uncover a wealth of substructure in planet-forming disks, including large, prominent spiral arms in MWC 758, SAO 206462, and V1247 Ori among others. In the classical theory of disk-planet interaction, these arms are consistent with Lindblad-resonance driving by multi-Jupiter-mass companions. Despite improving detection limits, evidence for such massive bodies in connection with spiral substructure has been inconclusive. In search of an alternative explanation, we use the PLUTO code to run 3D hydrodynamical simulations with two comparatively low planet masses (Saturn-mass, Jupiter-mass) and two thermodynamic prescriptions (three-temperature radiation hydrodynamics, and the more traditional $\beta$-cooling) in a low-mass disk. In the radiative cases, an $m = 2$ mode, potentially attributable to the interaction of stellar radiation with gap-edge asymmetries, creates an azimuthal pressure gradient, which in turn gives rise to prominent spiral arms \revision{in the upper layers of the disk}. Monte Carlo radiative transfer (MCRT) post-processing with RADMC3D reveals that \revision{in near-infrared scattered light, } these \revision{gap-edge} spirals are significantly more prominent than \revision{the} traditional Lindblad spirals \revision{for planets in the mass range tested}. Our results demonstrate that even intermediate-mass \textbf{proto}planets---less detectable, but more ubiquitous, than super-Jupiters---are capable of indirectly inducing large-scale spiral disk features, and underscore the importance of including radiation physics in efforts to reproduce observations.
  
}


   \maketitle
%
%-------------------------------------------------------------------

\section{Introduction}
In recent years, high-resolution near-infrared (NIR) imaging with instruments such as VLT/SPHERE, Gemini/GPI, and Subaru/HiCIAO has revealed large-scale spiral structure in a number of systems \citep[e.g.,][and references therein]{Shuai2022}, including MWC 758 \citep{Grady2013}, SAO 206462/HD 135344B \citep{Muto2012}, and V1247 Ori \citep{Ren2024}. %\dmnote{Mention also the ALMA system Elias 2-27, which has similar structure but not yet observed in scattered light.}
%[Several proposals exist to explain these structures, among them gravitational instability (GI) or flybys by external brown dwarfs or stars. However, the arms excited by GI move at the local epicyclic frequency and would be expected to ''roll up'' over several orbital periods, and so the open spirals we see represent only a very short-lived phase of the disk's evolution. Flybys, which would excite inner Lindblad resonances, are compatible with the constant pitch angles observed in these systems, but efforts to link systems hosting observed spirals to potential flyby candidates using GAIA position and velocity data have shown no correlation. While these mechanisms may be applicable on a case-by-case basis, they lack the applicability to explain the population of substructures as a whole.]
Given the ubiquity of planets in our Galaxy, disk-planet interaction has been suggested as a means of creating large-scale, multi-armed spiral structure in disks (among other proposals, including the gravitational instability and stellar flybys). Specifically, hydrodynamical simulations \citep[e.g.,][]{Fung2015,Zhu2015,Dong2016} of planetary wave excitation at the Lindblad resonances \citep{Goldreich1978,Goldreich1979,Goldreich1980} suggest that the observed scattered-light spirals are consistent with launching by exterior, multi-Jupiter-mass planets. Such companions, widely separated from their host stars \citep{vanderMarel2021}, should be detectable with angular differential imaging (ADI) in the near-infrared \citep{AsensioTorres2021, Desidera2021}, in the mid-infrared with the \textit{James Webb Space Telescope} (JWST) \citep[e.g.,][]{Wagner2019,Wagner2024,Cugno2024}, and in accretion tracers such as $H\alpha$ and $Pa\beta$ \citep{Follette2023,Biddle2024}, but evidence thus far is inconclusive. Only in the PDS 70 system are planets confirmed to exist \citep{Mueller2018,Keppler2018,Haffert2019} alongside disk substructure of any sort, with additional detections in e.g., AB Aur \citep{Currie2022} and HD 169142 \citep{Hammond2023}.

What, if not easily-detected, widely-separated super-Jupiters, might generate the observed spirals? Disk illumination and shadowing may have a role to play. \cite{Montesinos2016,Montesinos2018,Cuello2019} and \citep{Su2024} ran 2D, vertically-averaged hydrodynamical simulations of disks with parametrized cooling, including radial dark lanes in their temperature structures to model the effect of a misaligned, shadow-casting inner disk. The resulting strong azimuthal pressure gradient excites $m = 2$ spiral arms; MCRT post-processing, assuming a Gaussian vertical profile with a scale height governed by the 2D temperature, revealed that such arms would feature prominently in NIR. \cite{Nealon2020}, using SPH simulations with full radiative transfer, and \cite{Qian2024}, using a parametrized cooling prescription in a grid simulation, simulate the dynamical consequences of shadowing from an inner disk. Separately, in their 3D, grid-based simulations with flux-limited diffusion \cite[FLD; ][]{Levermore1981}, \cite{Chrenko2020} find that the outer edge of the gap carved by a $1 M_J$ planet exhibits unstable, non-axisymmetric flow patterns and additional spiral structure, which are absent in their locally-isothermal runs. They attribute this outcome to the fact that the temperature profile in the radiative case---with a cooler, shadowed gap region and warmer, directly illuminated outer gap edge---is more favorable for the excitation of the Rossby wave instability \citep[RWI; ][]{Lovelace1999}.

Taken together, these works suggest that even intermediate-mass planets could produce prominent, observable spirals, not through direct driving but by their impact on disk illumination---provided that shadow corotates with some radius within the disk. To investigate this hypothesis, we run 3D hydrodynamical simulations of disk-planet interaction with \texttt{PLUTO}, and post-process our results with the \texttt{RADMC3D} Monte Carlo radiative transfer code to generate mock near-infrared images of the simulated disk. In Section 2, we describe our methods in more detail, including our implementations of ``three-temperature'' radiation hydrodynamics \citep{Muley2023} and $\beta$-cooling to an axisymmetric background. In Section 3, we discuss our results, including the structure of the planet-formed gap, excitation of spirals at the gap edge, and post-processed near-infrared images. In Section 4 we present our conclusions and questions for future work.

\section{Methods}
\subsection{PLUTO radiation hydrodynamics}
For our study of spiral arms, we used a version of the \texttt{PLUTO} hydrodynamical code \citep{Mignone2007}, modified to solve the equations of radiation hydrodynamics \citep{MelonFuksman2019,MelonFuksman2021} with an additional dust internal energy field \citep{Muley2023}. This is thermally coupled to the gas- (via collision) and radiation-energy (via opacity) fields, but is a passive tracer of the gas velocity without any inertia or back-reaction (implying a Stokes number $\mathrm{St} \ll 1$ and globally constant dust-to-gas ratio $f_d \ll 1$). Radiation transport is computed using the M1 closure, which interpolates between the optically-thick (diffusion) and optically-thin (free-streaming) regimes \citep{Levermore1984}. We briefly review the most pertinent details below, while referring the reader to the aforementioned works (as well as \cite{Muley2024}) for a more detailed description of the equations and solution strategy.

\subsubsection{Three-temperature simulations}\label{sec:three_temp}
Our three-temperature module solves the full set of equations described in \cite{Muley2023}. The coupling of dust energy with radiation energy ($G_d$), and of dust momentum with radiation flux ($\vec{G}_d$), are given as follows:
\begin{subequations}
%\begin{equation}
%\begin{split}
%    G_g \equiv &-\rho \kappa_{g}(a_r T_g^4 - \xi_r) + r_g(\vec{\beta})\\
%\end{split}
%\end{equation}
\begin{equation}
\begin{split}
    G_d \equiv &-\rho \kappa_{d} f_d (a_r T_d^4 - E_r) + r_d(\vec{\beta})
\end{split}
\end{equation}
%\begin{equation}
%\begin{split}
%    \vec{G}_g \equiv & \rho \chi_g \vec{F}_r + \vec{r}_g(\vec{\beta})
%\end{split}
%\end{equation}
\begin{equation}
\begin{split}
    \vec{G}_d \equiv & \rho f_d \chi_d \vec{F}_r  + \vec{r}_d(\vec{\beta})
\end{split}
\end{equation}
\end{subequations}
where \revision{$\rho$ is the dust density, while $a_r = 4\sigma_{\rm Stefan-Boltzmann}/c$ is the radiation constant.} $T_d$ refers to dust temperature, while $E_r$ refers to the radiation energy density. $\kappa_d$ is a Planck-averaged (and thus, temperature-dependent) dust absorption opacity, while $\chi_d$ is the dust total (absorption plus scattering) opacity. In practice, we equate $\chi_d$ to the greater of the Planck- or Rosseland-averaged \textbf{dust absorption} opacities at a given temperature. $r_d(\vec{\beta})$ and $\vec{r}_d(\vec{\beta})$ are relativistic corrections to the above terms, with the vector $\vec{\beta} \equiv \vec{v}/c$ (not to be confused with the $\beta$-cooling timescale); these are negligible in the context of protoplanetary disks. In this work, we \revision{use the tabulated dust absorption opacities from \citep{Krieger2020,Krieger2022}, while setting} gas opacities to zero.

We also include a dust irradiation term, $S_{\rm irr}^d$, obtained by frequency-dependent ray-tracing from the central star. Here, the effective dust opacity is a Planck average taken at the stellar surface temperature $T_{*}$.

The dust-gas collision term $X_{gd}$ is defined as
\begin{equation}
    X_{gd} \equiv t_c^{-1}(r_{gd} \xi_d - \xi_g)
\end{equation}
where $t_c$ is the dust-gas thermal coupling time, $\xi_g$ and $\xi_d$ are the internal energy of dust and gas respectively, and $r_{gd} = c_g/ f_d c_d$ is the ratio of heat capacity per unit volume between gas and dust. $c_d$ is the specific heat capacity of dust, while $c_g \equiv k_B/\mu m_H (\gamma - 1)$ is that of the gas. The gas thermal coupling time $t_c$, as a function of the dust-gas stopping time $t_s$, is calculated in the Epstein regime \citep{BurkeHollenbach83,Speedie2022}. 

\begin{equation}
\label{eq:tcool}
    t_c = \frac{2/3}{\gamma - 1} f_d^{-1} t_s \eta^{-1}
\end{equation}
where we set the ``accommodation coefficient'' $\eta$ to unity. 

\subsubsection{$\beta$-cooling simulations}\label{sec:beta_cooling}
In our $\beta$-cooling simulations, we ignore absorption, emission, scattering ($G_d$, $\vec{G}_d$) and transport ($\vec{F}_r$ ) of thermalized radiation, as well as dust energy ($E_d$) and stellar irradiation ($S_d^{\rm irr}$). To parametrize all of these effects, we replace $X_{\rm gd}$ with a term of the form
\begin{equation}
    X_{\rm rel} \equiv t_{\rm rel}^{-1} \frac{\rho k_B}{\mu m_p (\gamma - 1)} \left(T_g - T_{\rm g, 0}(r, \theta)\right)
\end{equation}

where $T_{g, 0}$ is the (axisymmetric) initial condition for gas temperature (see Section \ref{sec:setup} for more details), $r, \theta, \phi$ are spherical coordinate positions, and $t_{\rm rel}$ can in general be a function of any primitive variables. Following \cite{Bae21} and \cite{MelonFuksman2023}, we set the thermal relaxation/cooling time
\begin{equation}\label{eq:t_rel}
    t_{\rm rel} = t_c + t_{\rm rad}
\end{equation}
where $t_c$ is defined as in Equation \ref{eq:tcool}, and $t_{\rm rad}$ is a radiative cooling timescale incorporating both the optically thick diffusion and optically thin free-streaming limits:
\begin{equation}
    t_{\rm rad} \equiv \max (\lambda_{\rm thin}^2, \lambda_{\rm diff}^2)/D
\end{equation}
where \revision{the thermal diffusivity} $D \equiv 4 c a_r T_g^3 / 3 c_g \kappa_R \rho^2$. The optically-thin cooling effective lengthscale $\lambda_{\rm thin} = (3 \kappa_R \kappa_P \rho^2)^{-1/2}$, while the thick diffusion length is taken to be the local scale height, $\lambda_{\rm diff} \equiv H = c_{s, \rm iso} \Omega^{-1}$, where $\Omega$ is the local Keplerian orbital frequency and $c_{s, \rm, iso} \equiv \sqrt{p/\rho}$ the ``isothermal sound speed.''

Cooling times were computed self-consistently in each cell of the grid \revision{at the start of the simulation}. For notational convenience, cooling times can be normalized by the local dynamical time $\Omega^{-1}$, with $\beta_{dg} = t_c \Omega$, $\beta_{\rm rad} = t_{\rm rad} \Omega$, and the overall $\beta = \beta_{\rm rad} + \beta_{\rm dg}$.

\begin{table}[]
\resizebox{\columnwidth}{!}{%
\begin{tabular}{l|l|l}
\hline
\multicolumn{1}{l|}{Symbol}             & \multicolumn{1}{l|}{Value}            & \multicolumn{1}{l}{Quantity}                                                        \\ \hline\hline
                                        &                                       &                                                                                      \\
$M_{*}$                                 & $1 M_{\odot}$                         & Stellar mass                                                                         \\
$R_*$                                   & $2.08 R_{\odot}$                      & Stellar radius                                                                       \\
$T_*$                                   & 4000 K                                & Stellar surface temperature                                                          \\
$\alpha$                                & 0.001                                 & \begin{tabular}[c]{@{}l@{}}Shakura-Sunyaev \\ viscosity parameter\end{tabular}       \\
$\mu$                                   & 2.3                                   & Mean molecular weight of gas                                                         \\
$\gamma$                                & 1.41                                  & Adiabatic index of gas                                                               \\
$c_d$                                   & $0.7 \si{\joule\per\gram\per\kelvin}$ & Specific heat capacity of dust                                                       \\
$c_g$                                   & $8.8 \si{\joule\per\gram\per\kelvin}$ & Specific heat capacity of gas                                                        \\
$\{n_r, n_\theta, n_\phi\}$    & ${134, 58,460}$  & \begin{tabular}[c]{@{}l@{}}Cells along each dimension \\ (low-resolution, 1000 orbits)\end{tabular} \\
$\{n'_r, n'_\theta, n'_\phi\}$ & ${268, 116,919}$ & \begin{tabular}[c]{@{}l@{}}Cells along each dimension \\ (high-resolution, 10 orbits)\end{tabular}  \\
$\{r_{\rm in}, r_{\rm out}\}$           & $\{{\rm 16 au, 100 au}\}$             & Inner and outer radial boundaries                                                    \\
$\{\theta_{\rm low}, \theta_{\rm up}\}$ & $\{-0.4, 0.4\} + \pi/2$                       & Lower and upper polar boundaries                                                     \\
$r_{\rm inf}$                           & 18 au                                 & \begin{tabular}[c]{@{}l@{}}Outer boundary of inner \\ wave-damping zone\end{tabular} \\
$r_{\rm sup}$                           & 95 au                                 & \begin{tabular}[c]{@{}l@{}}Inner boundary of outer\\ wave-damping zone\end{tabular}  \\
$r_p$                                   & 40 au                                 & Planet radius                                                                        \\
$\Omega_p^{-1}$                         & 40 y                                  & Dynamical time at $r_p$                                                              \\
                                        & $3 \times 10^{-4} M_{\odot}$          & Mass of Saturn                                                                       \\
                                        & $1 \times 10^{-3} M_{\odot}$          & Mass of Jupiter                                                                      \\
                                        &                                       &                                                                                     
\end{tabular}%
}
\caption{Table of key simulation parameters. More information can be found in Section \ref{sec:setup} of this work, and in Section 2 of \cite{Muley2024}.}
\end{table}

\subsubsection{Setup and initial conditions}\label{sec:setup}
\begin{figure}
    \centering
    \includegraphics[width=0.5\textwidth]{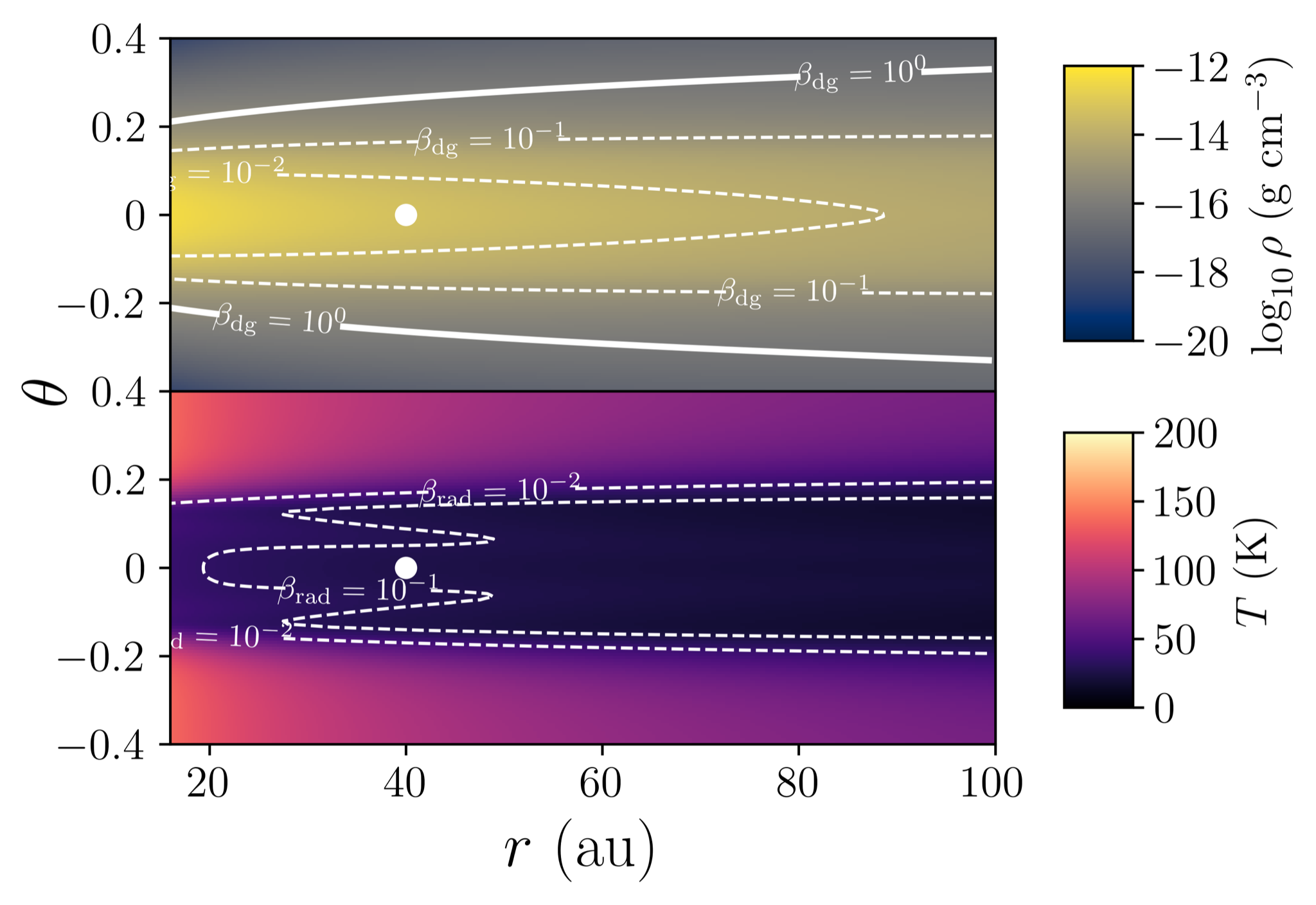}
    \caption{Initial conditions for gas density $\rho$ (above), \revision{as well as dust and gas temperatures $T_d$ and $T_g$ (below), which are initially equal.} Contours for normalized cooling timescales from gas-grain collision ($\beta_{\rm dg}$) are plotted above, and from radiation ($\beta_{\rm rad}$) below, with white lines. Figure reproduced exactly from \cite{Muley2024}.}
    \label{fig:initial_conditions}
\end{figure}
For our simulations, we used the same initial conditions (plotted in Figure \ref{fig:initial_conditions}) as in \cite{Muley2024}, generated with iterative alternation of radiative and hydrostatic computations. As in that work, we use an extended radial grid with \revision{an} inner edge at 0.4 au, as well as an additional inner padding prescription, to ensure an accurate temperature at the inner edge of our simulated domain. Dust 
%and stellar 
properties remain unchanged between that work and the present one, and dust temperature $T_d$ and gas temperature $T_g$ are assumed to be equal in the initial condition. The specific heat capacity for dust, $c_d = 0.7 \ \si{\joule\per\gram\per\kelvin}$.

The disk's surface density
\begin{equation}
    \Sigma_g = 200 \ \si{\gram\per\cm\squared} \left(\frac{R}{1 {\rm au}}\right)^{-1} \,
\end{equation}

where $R = r \sin \theta$ is the cylindrical radius. \revision{We use a \citep{Shakura1973} $\alpha = 0.001$.}

We fix the stellar mass $M_* = 1 \ M_{\odot}$, radius $R_* = 2.08 \ R_{\odot}$, and temperature $T_* = 4000$ K, giving a total luminosity equal to solar. We test two thermodynamic prescriptions ($\beta$-cooling, 3T) and planet masses ($M_p = 3 \times 10^{-4} M_{\odot}, 1 \times 10^{-3} M_{\odot}$), with the planet located at $r_p = 40 {\rm au}, \theta_p = \pi/2$, and $\phi_p = \pi/4$ in each case. \revision{The planets are introduced relatively rapidly into the simulations, with a growth timescale $t_{\rm gr} = 10 y$; this has the potential to strengthen gap-edge effects such as the Rossby wave instability \citep{Hammer2017}, and thus serves as a caveat of our model.} \revision{For our hydrodynamic computations, we take a sub}domain \revision{of the initial disk model generated by our hydrostatic procedure, bounded by} $r = \{0.4, 2.5\} \times r_p$, $\theta = \{-0.4, 0.4\} + \pi/2$, and $\phi = \{0, 2\pi\}$. We cover this box with $134 (r) \times 58 (\theta) \times 460 (\phi)$ cells, logarithmic in radius and uniform in polar and azimuthal angle. This yields a resolution of ${\sim} 4.5$ cells per scale height at the planet location, roughly equal to that used in the radiation-hydrodynamic gap-opening simulations of \cite{Chrenko2020}.  

We fix the boundary values of all fields to equal that of the initial condition. \revision{To ensure numerical stability near these boundaries, we damp $\rho$, $E_r$, $S_d^{irr}$, and the components of $\vec{v}$ to their initial conditions over a timescale  $t_{\rm damp} = 2\pi \times 0.1 \Omega^{-1}$.} We use an HLL Riemann solver for hydrodynamics and HLLC for radiation; to ensure numerical stability in the presence of fast flows and large density contrasts, we make use of a piecewise-linear method (PLM) reconstruction to obtain the states at cell boundaries, rather than the WENO3 method used in \cite{Muley2024}. We run each simulation for $250 000 {\rm yr}$ (approximately 1000 planetary orbits), then interpolate the result onto a higher-resolution $268 (r) \times 116 (\theta) \times 919 (\phi)$ grid; we run this setup for an additional $2500 {\rm yr}$ to study flow patterns in more detail.

\subsection{RADMC3D Monte Carlo radiative transfer}\label{sec:radmc3d}
We used \texttt{RADMC3D} \citep{Dullemond2012} to generate \revision{VLT/SPHERE} \textit{H}-band ($\lambda_H = 1.62 \mu m$) images from our high-resolution simulation outputs, with $N_{\rm phot} = 2.5 \times 10^7$ photon packages launched from the central star. In post-processing of 3T models, dust temperature was taken directly from the simulation output, whereas for $\beta$-cooling, it is assumed equal to the gas temperature. To minimize spurious illumination at the inner edge of our hydrodynamic simulation domain, located at $r_{\rm in} = 16 {\rm \ au} $, we pad the disk to $r_{\rm pad} = 0.5 {\rm \ au}$ using the initial conditions generated in Section \ref{sec:setup}. 
\\
\revision{For simplicity and consistency in our radiative-transfer calculations, we used the tabulated dust opacities in \citep{Krieger2020,Krieger2022}, including absorption, scattering, and \cite{HenyeyGreenstein1941} $g$ parameters to accommodate directional scattering;} we expect \revision{that a more comprehensive calculation including polarization} would yield similar results. In generating our \revision{$H$-band} images, we assume a fiducial line-of-sight distance $d = 100 {\rm pc}$, and convolve them with a Gaussian kernel (${\rm FWHM} \approx 2.35\sigma = 0.06 {\rm arcsec}$) to mimic the effects of instrumental beam size. 

\section{Results}

\begin{figure}
    \centering
    \includegraphics[width=0.5\textwidth]{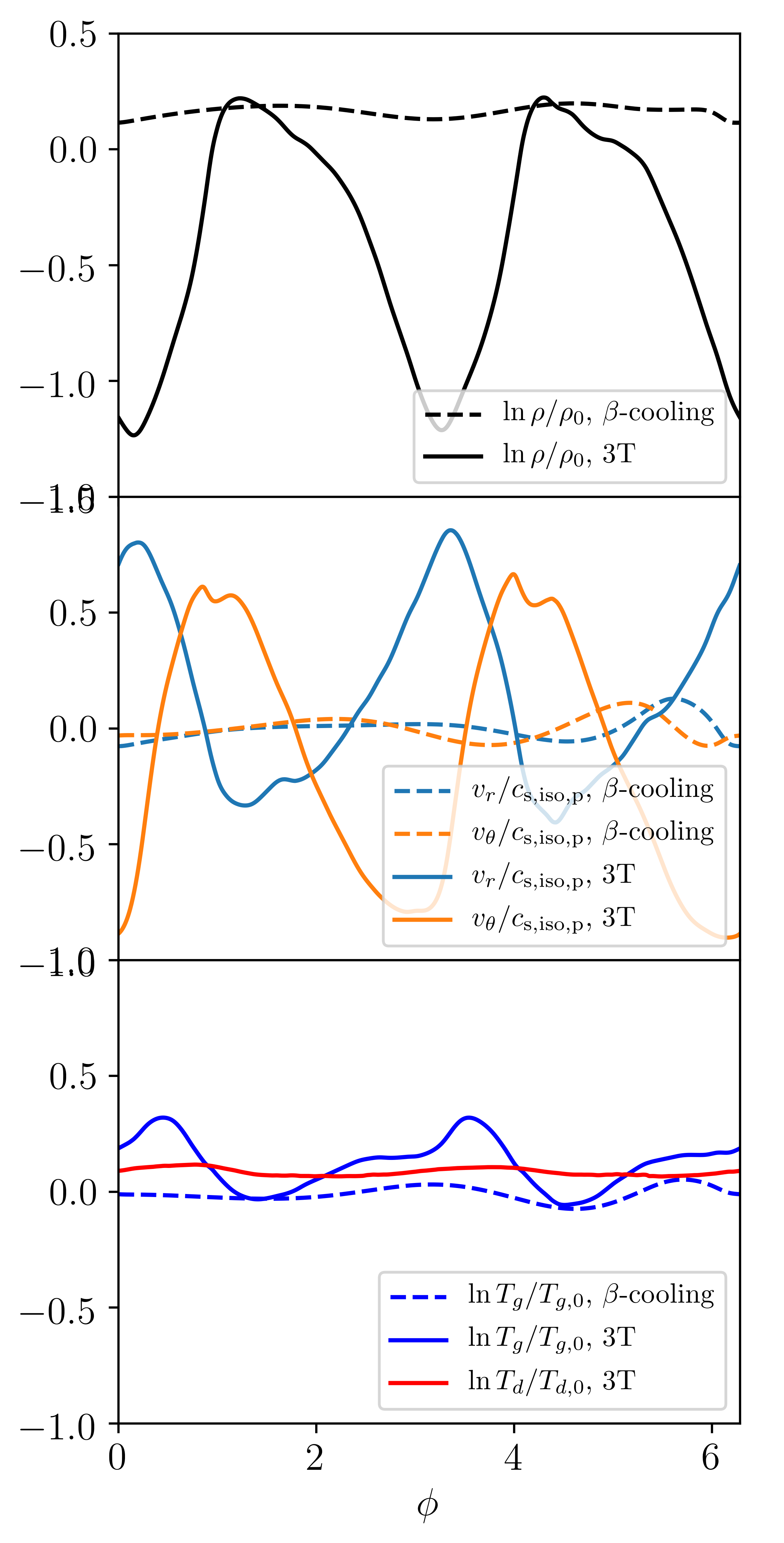}
    \caption{Azimuthal profiles of various quantities at a fiducial radius of $r = 60 {\rm \ au} = 1.5 r_p$, and at an altitude of $\theta = 0.2$ rad above the midplane \revision{in our $M_p = 3 \times 10^{-4} M_{\odot}$ simulations}. In the 3T case, azimuthal asymmetries of even a few percent in disk illumination---potentially induced by the RWI, and visible in the $T_d$ profile---lead to strong non-axisymmetry in $\rho$, $T_g$, and the velocity components.
    }
    \label{fig:azi_profiles}
\end{figure}

\begin{figure*}\label{fig:dens-compare}
    \centering
    \includegraphics[width=0.99\textwidth]{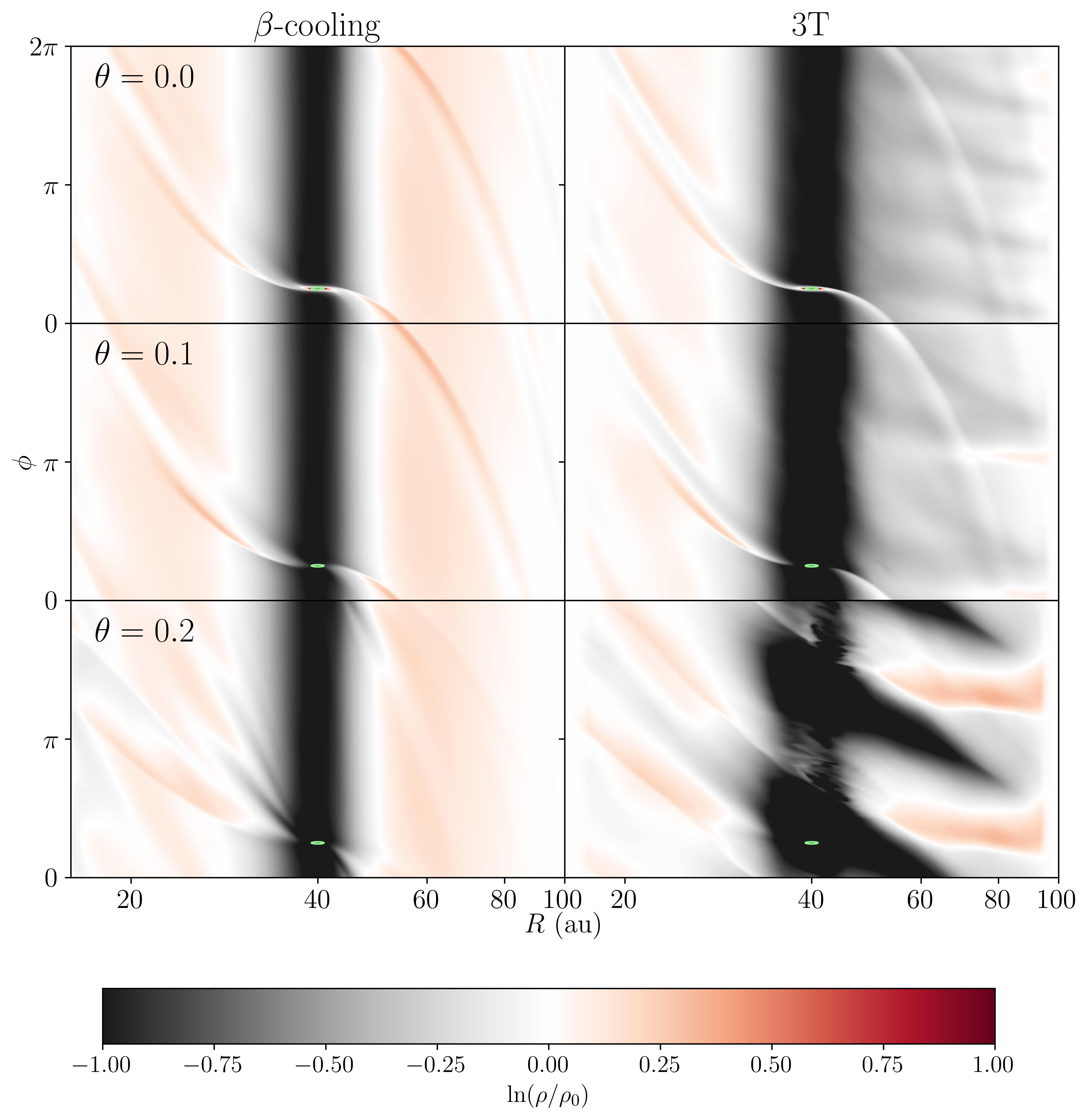}
    \caption{Gas density $\rho$ at $t = 1010$ orbits (1000 orbits at the fiducial resolution, 10 at the doubled resolution used in \cite{Muley2024}), with respect to the initial condition $\rho_0$, for our simulations with a Saturn-mass planet. The green ellipse indicates the planet's Hill radius. With the 3T scheme, the disk atmosphere shows clear development of $m = 2$ spiral arms, which are absent in the $\beta$-cooling simulations. \revision{White bands at the interior and exterior of the radiative simulation result from wave-damping to the initial condition.}
    }
\end{figure*}

\subsection{Temperature structure and thermal spirals}\label{sec:spiral-temp}
During our simulations, the Lindblad spirals excited by the planet increased the angular momentum of the disk material exterior to the planet's orbit and decreased that of the material interior; the cumulative effect of this, over many orbits, is to open a gap in the disk. Once the gap became sufficiently deep---at roughly 50 orbits for the Jupiter-mass planet, and 200 for the Saturn-mass planet---the outer gap edge deviates from axisymmetry, potentially due to hydrodynamic instabilities such as the Rossby wave instability \citep[see Section 3.1.6 of][for a more detailed discussion]{Chrenko2020}.
\revision{With} $\beta$-cooling, a viscous $\alpha = 1 \times 10^{-3}$ and an axisymmetric background temperature rapidly damp azimuthally-asymmetric structures to restore the classical picture of disk-planet interaction \citep{Zhang2024}. With full radiation hydrodynamics, however, the density asymmetries at the outer gap edge couple with the stellar radiation field to create asymmetries in the temperature structure as well, which in turn alters the density, pressure, velocity, and vorticity (Appendix \ref{sec:vorticity}). Although the vertically-integrated temperature deviation is fairly small---typically no more than 10\%---the perturbations in the disk atmosphere became substantial, as disk material adjusted its scale height over the course of an orbit \revision{and had work performed on it due to compression and expansion (via the $P dV$ term). In this region---which we define to lie at or above the radial $\tau_r = 1$ surface for stellar irradiation---gas thermal relaxation is dominated by the gas-grain cooling time $t_c \approx 0.1-1 \Omega^{-1}$, ensuring that this $P dV$ heating was not immediately radiated away.}

\revision{The temperature asymmetry leads to a pressure asymmetry at the outer gap edge, which drives strong spiral arms \cite[e.g.,]{Montesinos2016,Cuello2019};} in our case, the $m = 2$ mode predominates. Contributing to the observed non-axisymmetry in the upper disk layers is the breakdown of midplane symmetry, with a net $|v'_{\theta}| \lesssim 0.05 c_{\rm s, iso, p}$ in the midplane, but significantly faster in the upper \revision{layers of the disk.} In Figure \ref{fig:azi_profiles}, we plot the resulting azimuthal profiles of density, temperature, and velocity components at high altitude near the outer gap edge \revision{at $t = 252 500$ y (${\sim}1000$ orbits at low resolution, plus ${\sim}10$ at the doubled resolution). In the bottom panel, the decoupling between dust and gas temperatures in 3T---mentioned in the previous paragraph---is clearly visible.} \revision{To better understand this phenomenon, we also conducted a controlled numerical experiment (Appendix \ref{sec:beta-spiral-excitation}) using $\beta$-cooling. We initialized the background temperature profile with asymmetries in azimuth and about the midplane. In this test, we observed the development of strong spiral structure in the disk atmosphere---as in the 3T fiducial simulation, where the asymmetric temperature structure developed naturally from illumination, rather than being prescribed}. 

In Figure \ref{fig:dens-compare}, we plot \revision{2D $r-\phi$ cuts of} the density structure at various \revision{polar} altitudes, comparing the results obtained with the three-temperature and $\beta$-cooling approaches. \revision{In both cases, the Lindblad spirals generated by planetary forcing are most prominent at $\theta = 0.0$ and $\theta = 0.1$ above the midplane, but weaken at $\theta = 0.2$ in the disk atmosphere. At these high altitudes, spirals excited by the pressure gradient generated by shadowing---with a somewhat larger corotation radius---become more prominent. Both sets of spiral waves propagate through the disk according to the WKBJ dispersion relation for the radial wavenumber $k_r$ and azimuthal wavenumber $m$ \citep[see][and references therein]{Bae2018a}
\begin{equation}
    m(\Omega - \Omega_c)^2 = \Omega^2 + k_r^2 c_s^2
\end{equation}
in which we assume that the epicyclic frequency $\mathcal{K} = \Omega$. From this, the pitch angle $\delta = \arctan(m/k_r r)$ can be derived as
\begin{equation}
    \delta = \arctan\left[\frac{c_s}{R \Omega(R)}\left(\left(1 - \Omega_c/\Omega(R)\right)^2 - m^{-2}\right)^{-1/2}\right] 
\end{equation}
where $\Omega_c$ is the corotation frequency of the perturbation and $m$ is the azimuthal wavenumber. The larger corotation radius for shadow-driven spirals accounts for the difference in pitch angles between the spirals, and mimics the effect of an external planetary driver. 

Over hundreds of orbits, the strong vertical flows shown in Figure \ref{fig:azi_profiles} have the effect of depleting the outer disk; this, in turn, would reduce the strength of the (inward-directed) outer Lindblad torque on the planet, slowing or even reversing inward Type II planetary migration. However, the precise rate of mass depletion from the domain depends on the strength of vertical damping (see also Appendix \ref{sec:boundaries} for a discussion on how the spiral strength is affected). Moreover, our simulations do not incorporate other sources of mass depletion which become relevant at high altitude, such as photoevaporation and disk winds. As such, we choose to remain agnostic about the true mass-loss rate, and defer such an investigation to future work.}

Despite sharing similarities with the models presented in \cite{Chrenko2020}, our simulations do have important differences. In contrast to our 3T, M1 numerical approach, they employ two-temperature (2T) FLD \citep[e.g.,][]{Bitsch2013a}, which cannot account for free-streaming and shadowing effects below the disk's radial $\tau = 1$ surface, or for the delayed response of gas to changes in illumination due to the nonzero collisional coupling time. They also only simulate the upper half of the grid, locking the disk to midplane symmetry, and impose strong damping of $v_\theta$ at the upper polar boundary to more strictly conserve mass within the grid (we find in \revision{Appendix} \ref{sec:boundaries} that this somewhat strengthens the observed spiral signature). These differences are a consequence of the separate, but complementary, research goals of each work---their emphasis is on gap structure and planetary torque and migration, while ours is on observational signatures in the upper disk atmosphere.

\subsection{Radiative transfer post-processing}
\begin{figure*}
    \centering
    \includegraphics[width=1\textwidth]{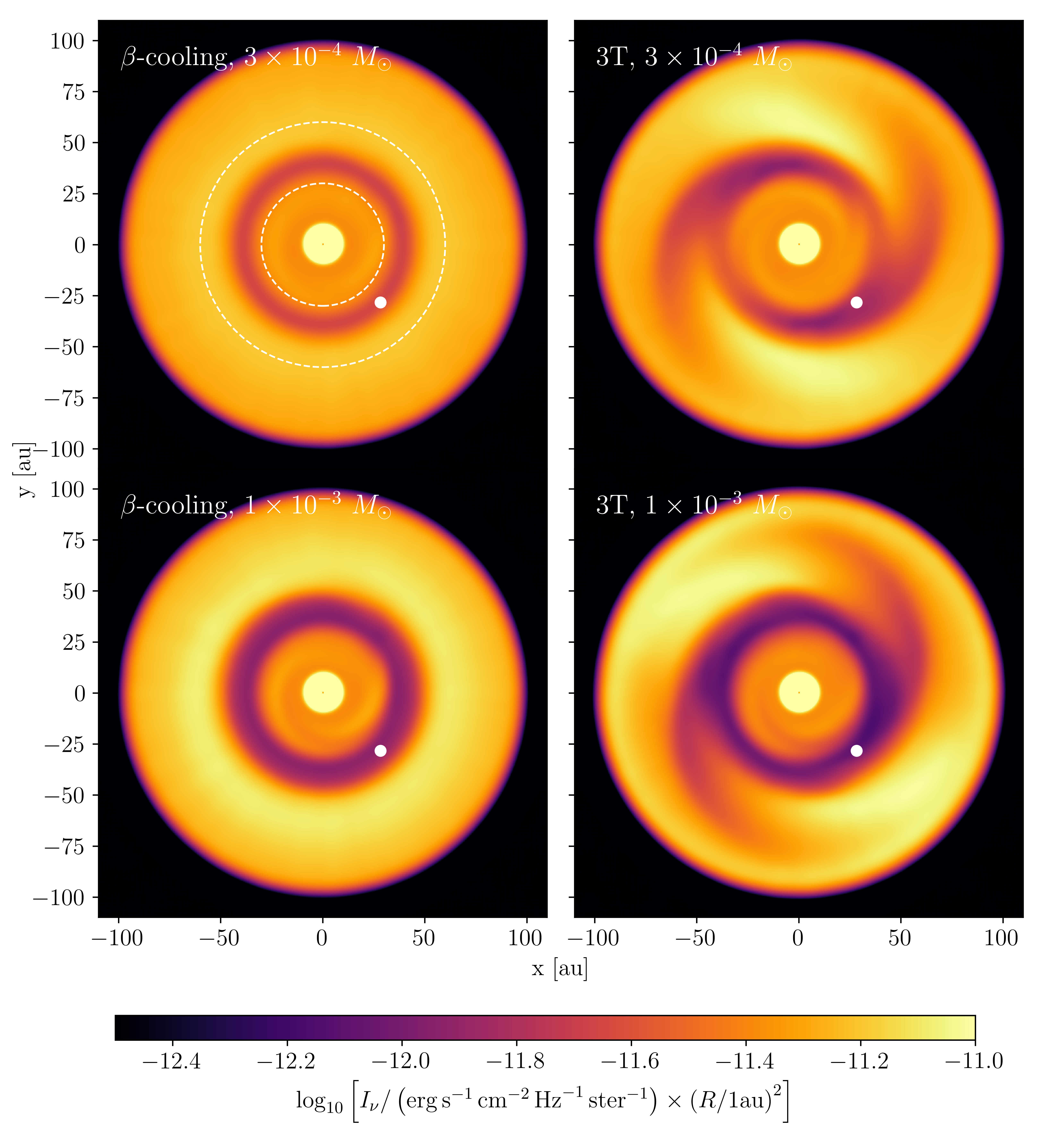}
    \caption{Mock $H$-band ($\lambda_H = 1.62 \mu m$) total intensity for the disks we simulate, including an inverse-square correction factor. We assume the disk is located at a line-of-sight distance $d_r = 100$ pc with a face-on orientation ($i_d = 0^{\circ}$), and convolve the raw image with a Gaussian of FWHM = 0.06 arcsec to mimic the effects of finite telescope resolution. The differences between the radiative and non-radiative cases are clearly visible. }
    \label{fig:NIR_mock_images}
\end{figure*}

\begin{figure}
    \centering
    \includegraphics[width=0.5\textwidth]{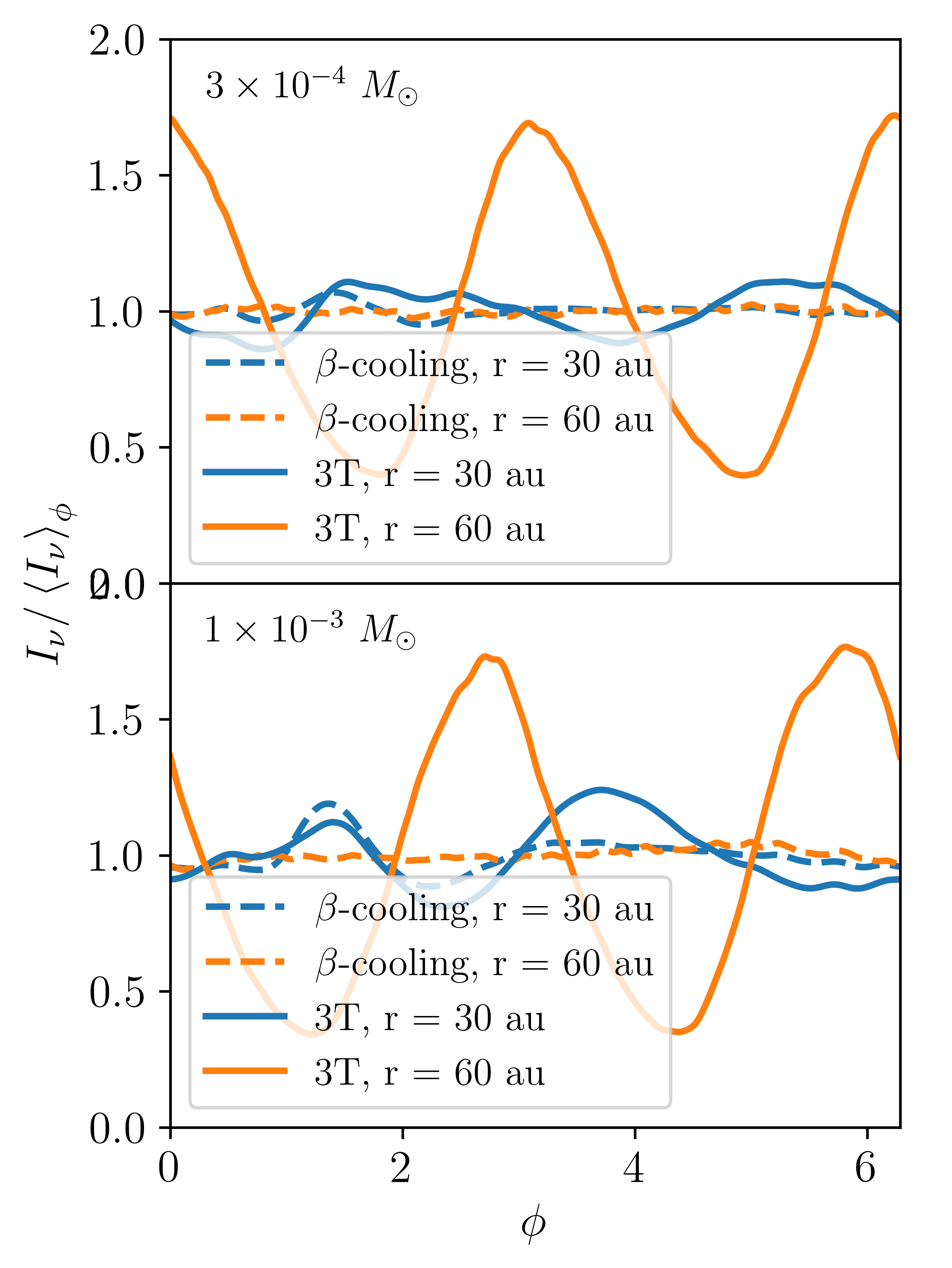}
    \caption{$H$-band intensity in all of our simulations, taken at fiducial radii $r_{\rm inner, cut} = 30$ au and $r_{\rm outer, cut} = 60$ au, and normalized to the azimuthal average. The 3T simulations show a clear $m = 2$ spiral with a peak-to-trough intensity ratio ${\sim} 4$ in the outer disk and ${\sim}1.2-1.5$ in the inner disk. By contrast, the $\beta$-cooling simulations show negligible asymmetry in the outer disk, alongside asymmetries of ${\sim}1.1-1.5$ in the inner disk caused by inner Lindblad spirals.}
    \label{fig:contrast}
\end{figure}

In Figure \ref{fig:NIR_mock_images}, we display face-on simulated NIR images generated with \texttt{RADMC3D} using the procedure described in Section \ref{sec:radmc3d}; in order to better show the spiral structure throughout the disk, the intensity is scaled by $(R/1 {\rm \ au})^2$. For the 3T models, the gap-induced spirals are prominently visible in NIR, far outweighing the planet-driven Lindblad spiral \revision{, whose contrast is diluted by beam convolution}. By contrast, the $\beta$-cooling simulations contain only the Lindblad spiral, which is greatly weakened by the effects of beam convolution.

For a more quantitative view, we plot 1D azimuthal profiles of the brightness contrast at selected radii in Figure \ref{fig:contrast}. In the outer disk (specifically, $r_{\rm outer, cut} = 60$ au), the peak-to-trough brightness contrast across the spiral reaches a factor of $\sim 4$ in 3T models, but is negligible for $\beta$-cooling models. In the inner disk ($r_{\rm inner, cut} = 30$ au), the 3T simulations show a contrast of 1.2-1.5, with larger values for higher planet mass. With $\beta$-cooling the perturbation amplitude is roughly the same, but results from the Lindblad spiral and not from any azimuthal temperature difference.

\section{Conclusions}
We conducted three-dimensional hydrodynamical simulations of disk-planet interaction, varying both planet mass (Saturn-mass, Jupiter-mass) and thermodynamic prescription (three-temperature radiation hydrodynamics, $\beta$-cooling). In the radiative case, deviations from axi- and midplane symmetry at the outer gap edge of a planet-carved gap are amplified by interaction with the radiation field. The resulting strong azimuthal pressure gradients in the upper disk launch spiral density waves \citep{Montesinos2016} with an $m = 2$ azimuthal mode structure. Mock $H$-band images, generated using the \texttt{RADMC3D} Monte Carlo radiative transfer code, showed azimuthal brightness contrasts of a factor of a few, \revision{equivalent to those from classical Lindblad spirals generated by multi-Jupiter-mass planets exterior to the spiral \citep{Dong2017}.} This radiative mechanism neither requires the existence \revision{of such a massive exterior companion}, 
%stronger than that associated with the classical Lindblad spirals generated by such intermediate-mass planets. \revision{\dmnote{Discuss Dong and Fung 2017 findings about spiral strength and comparison to Lindblad spirals.}} This radiative mechanism neither requires the existence of a multi-Jupiter mass planet exterior to the spiral, 
nor creates a correspondingly strong spiral structure in the disk midplane which would concentrate millimeter grains. This is consistent with multi-wavelength observations of systems such as V1247 Ori \citep{Ren2024}.

Although our simulations include only Saturn- and Jupiter-mass planets, the spiral-excitation mechanism described here would likewise operate at the outer edge of a gap carved by a super-Jupiter companion. Being interior to the spiral and close to its host star on the sky, such a body would be more easily concealed by the stellar point-spread function than an exterior companion required by the inner-Lindblad-spiral hypothesis. Indeed, high-mass interior companions may provide a natural explanation for observations in systems such as MWC 758 \citep{Boehler2018,Dong2018} and SAO 206462 \citep{vanderMarel2016}, where deep \citep{FSC14}, eccentric \citep{Goldreich2003,Kley2006} cavities in millimeter continuum and gas tracers exist alongside large-scale NIR spirals.

\revision{In principle, the instability we observed could have been obtained using an orthodox, two-temperature approach. However, with 2T, the gas temperature (and consequently, its vertical structure) responds immediately to changes in illumination, which contributes contributes to the development of azimuthal asymmetries at the inner boundary of the disk (also found in Renggli and Szulágyi 2022, private communication). These features, in turn, affect the temperature structure and generate spiral arms even in the absence of a planet. With 3T, however, the nonzero coupling time between gas and grains suppresses these inner-boundary artifacts, leaving only the spirals generated at the outer gap edge.}

With the existing 3T implementation in \texttt{PLUTO}, one could conduct a parameter study over disk/stellar mass, planet location, and opacity prescription (including molecular opacity) to better understand the properties that give rise to large-scale spirals \citep{vanderMarel2021}, and test whether the mechanism we describe can form spiral structures with wavenumbers besides $m = 2$. Line transfer, a feature of \texttt{RADMC3D}, could be used to model the kinematic signatures expected from such large-scale spirals, and which would potentially be observable in exoALMA and similar large programs. Incorporating multiple dust species  \cite[e.g.,][]{Krapp2024}, would allow the dynamics of small and large dust grains to be handled simultaneously, ensuring that post-processed near-infrared and submillimeter images are mutually consistent. Using radiation-transport methods that allow for beam crossing \citep[such as the discrete-ordinates approach of ][]{Jiang2021} would improve the treatment of shadowing, particularly in the presence of an accreting, luminous planet that radiates in all directions. Pursuing these potential future directions would greatly expand scientific understanding of how radiation impacts disk-planet interaction, and in doing so, help reproduce observations of real systems.

\begin{acknowledgements}
We thank Marcelo Barraza Alfaro, Jiaqing Bi, Ondřej Chrenko, Gabriele Cugno, Ruobing Dong, Mario Flock, \revision{Anton Krieger}, Nienke van der Marel\revision{, Jacksen Narvaez}, Rebecca Nealon, Yansong Qian, Richard Teague,  Bhargav Vaidya, and Alexandros Ziampras for useful discussions. \revision{We thank the anonymous referee for a thorough and insightful report which helped us improve the quality of this work.} Numerical simulations were run on the Cobra and Raven clusters of the Max-Planck-Gesellschaft and the Vera Cluster of the Max-Planck-Institut f\"ur Astronomie, both hosted by the Max Planck Computing and Data Facility (MPCDF) in Garching bei München. The research of J.D.M.F. and H.K. is supported by the German Science Foundation (DFG) under the priority program SPP 1992: ``Exoplanet Diversity" under contract KL 1469/16-1/2. 

\end{acknowledgements}

%\nocite{*}
\bibliography{threetemp_spiral_aanda}
\bibliographystyle{aa}
%\begin{comment}
\appendix
\section{Vorticity of the flow field}\label{sec:vorticity}
\noindent\begin{minipage}{\textwidth}
    \centering
    \includegraphics[width=1\textwidth]{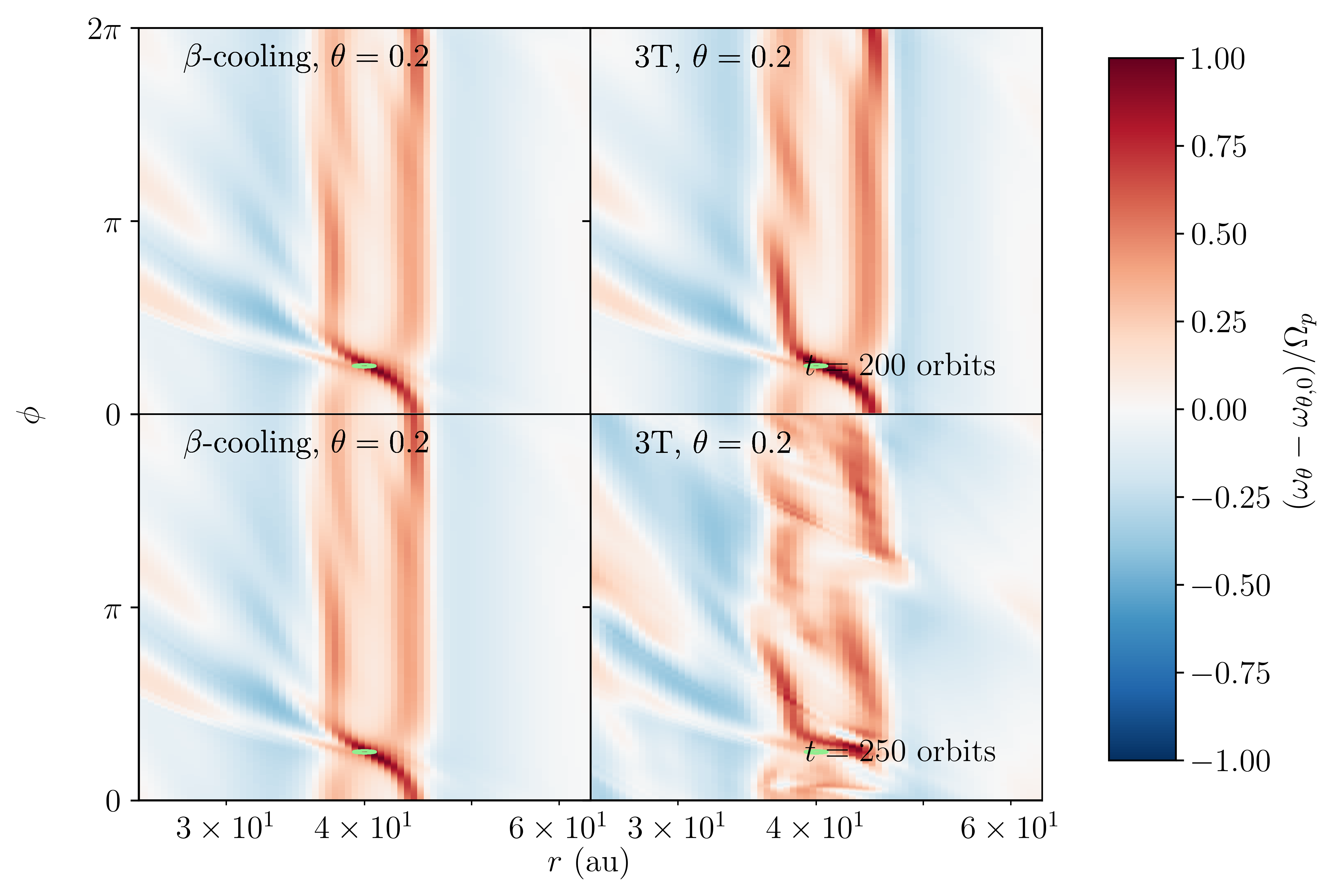}
    \captionof{figure}{Plots of the $\theta$-component of vorticity at $\theta=0.2$, with respect to the initial condition, in units of the planet's orbital frequency $\Omega_p$. Between 200 and 250 orbits, the vorticity profile is largely unchanged in the $\beta$-cooling simulations, but deviates strongly in the 3T case as radiation couples to the asymmetries generated by disk-planet interaction. }
    \vspace{24pt}
    \label{fig:vorticity_final}
\end{minipage}
The vorticity provides a complementary view to that offered by Figure \ref{fig:dens-compare}, emphasizing the gap itself rather than the spiral arms. In Figure \ref{fig:vorticity_final}, we plot the polar component of vorticity, $\omega_\theta = (\nabla \times \vec{v}) \cdot \hat{\theta}$, for our simulations with $M_p = 3 \times 10^{-4} M_\odot$, comparing the $\beta$-cooling and 3T prescriptions side-by-side. At $t = 200 $ orbits, right at the onset of the instability described in \ref{sec:spiral-temp}, the vorticity profile is similar between simulations. At $t = 250$ orbits, the gap profile with $\beta$-cooling remains intact, but that with 3T is substantially disrupted at the outer edge, causing spiral density waves to be launched. 

With 3T, the gap profile bears some resemblance to that caused by the irradiation instability of \cite{Fung2014}. However, the instability that they study is a consequence of radiation pressure directly influencing gas momentum, rather than radiative heating altering the disk's vertical structure. In this latter respect, the simulated mechanism can be analogized to the self-shadowing instability of \cite{Wu2021}. 

\FloatBarrier
\clearpage
\section{$\beta$-cooling tests}\label{sec:beta-spiral-excitation}
\begin{figure}
    \centering
    \includegraphics[width=0.5\textwidth]{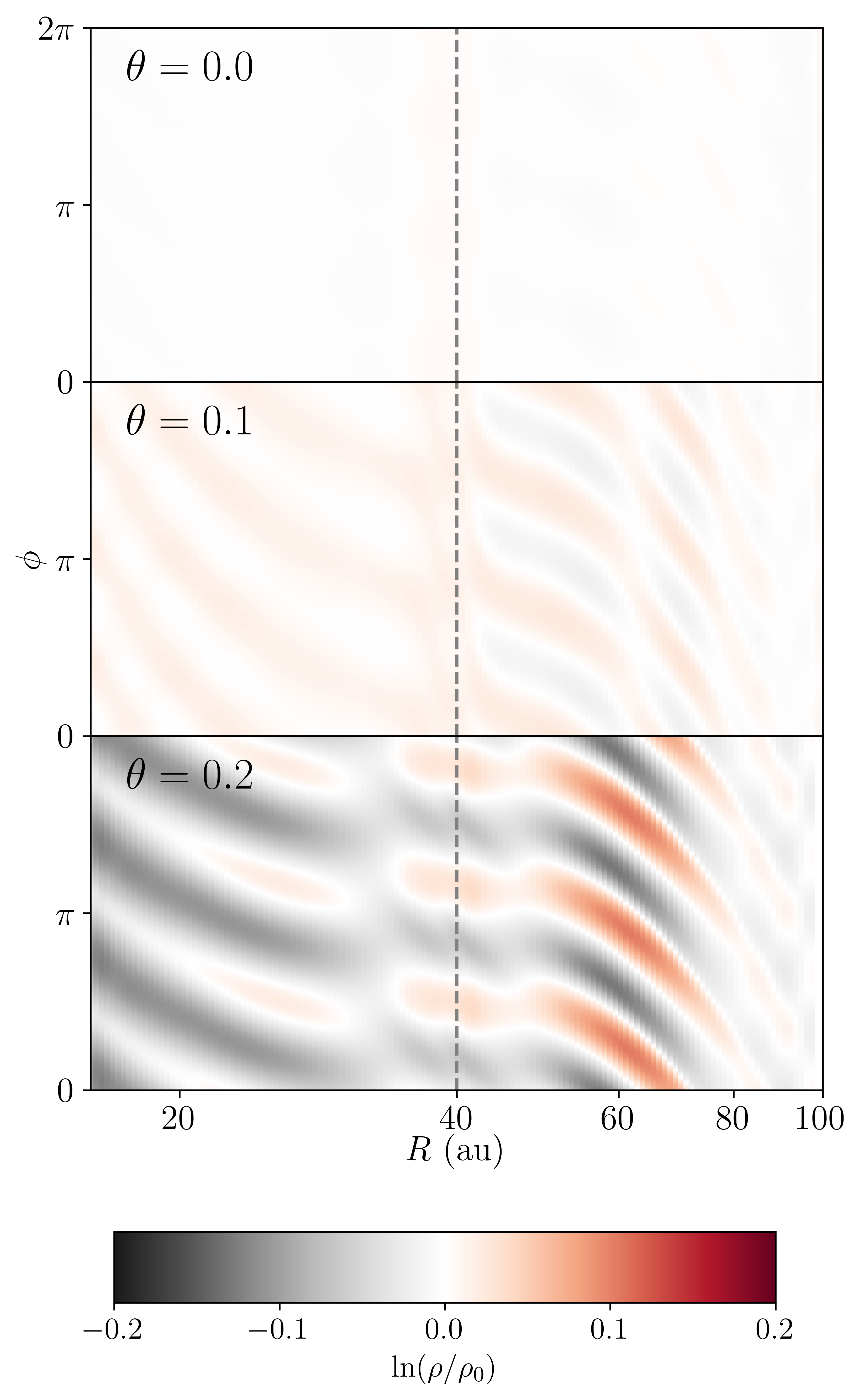}
    \caption{Plot of density perturbation in special $\beta$-cooling simulations where the planet is removed, and for which the temperature profile in Figure \ref{fig:initial_conditions} is multiplied by the azimuthal perturbation factor $\mathcal{W}_T(\phi, t)$ (Equation \ref{eq:temp_factor}). The corotation radius of the temperature profile, equal to the planet radius in the main text, is plotted as a dotted vertical line. Note that the density scale has a much smaller range than in Figure \ref{fig:dens-compare} in the main text. }
    \label{fig:betacool-temp}
\end{figure}

To validate our hypothesized mechanism for the generation of spirals, we ran $\beta$-cooling simulations with no planet. We used the same background temperature profile as in the main text, but multiplied it everywhere by an azimuthally varying factor of
\begin{equation}\label{eq:temp_factor}
    \mathcal{W}_T(\phi, t) = 1 + f_t \sin (\theta) \sin (m(\phi - \Omega_p t)) \exp \left(-(r/r_T)^{-4}\right)
\end{equation}
where $f_t = 0.25$, $\Omega_p$ is the Keplerian orbital frequency at $r_p = 40 $au, and $r_T = 32$ au. We choose an $m = 3$, as opposed to the $m = 2$ the observed system settles into, to show the full generality of the mechanism. 
Although the temperature deviation is at most 10\% within the domain, the effect it has on the disk's vertical structure means that the spiral signature becomes quite substantial in the upper atmosphere. 

In Figure \ref{fig:betacool-temp}, we plot the resulting density profile at $t = 40$ orbits, by which the system has settled into a quasi-steady state. Because the corotation radius of the temperature pattern ($r_p = 40 $au) is closer than the outer gap edge in the 3T simulations, the expected outer set of spirals is also contained within the domain. In our self-consistent radiation-hydrodynamics runs, the RWI mediates spiral formation insofar as it creates an analogous non-axisymmetric temperature structure, rather than through direct launching \citep[as in, e.g., ][]{Huang2019}.
\section{Boundary conditions}\label{sec:boundaries}
\begin{figure}
    \centering
    \includegraphics[width=0.5\textwidth]{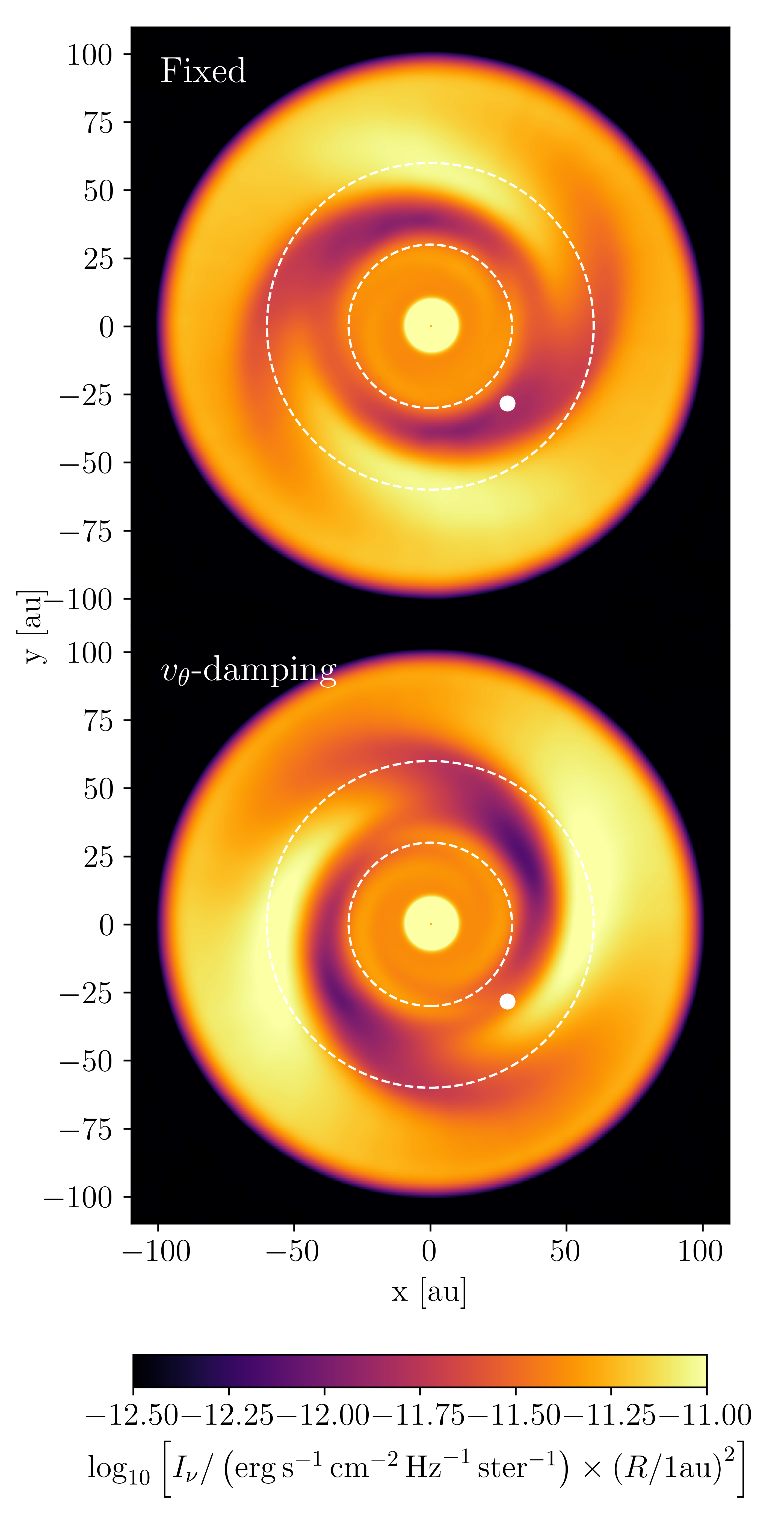}
    \caption{Mock $H$-band images for different wave-damping prescriptions at the boundary.}
    \label{fig:h_band_bc}
\end{figure}
Radiation-induced spirals---absent in our $\beta$-cooling simulations with axisymmetric background temperatures---induce strong flows in the vertical direction (formally, $v_{\theta}$) in the upper layers of the disk. These, together with with our fixed condition at the upper and lower $\theta$-boundaries of the domain, allow mass to escape from the simulation domain over long periods of time. This complicates radiation-hyrodynamical studies of planetary torques, migration, and transition-disk carving, for which an accurate surface density profile is essential.

Mass can be better conserved within the domain by strongly damping $v_{\theta}$ to zero in the uppermost and lowermost regions of the simulation domain \citep[see also][who damped $v_{\theta}$ to an azimuthal average]{Chrenko2020}. However, this would inevitably lead to wave reflection, which could affect the development of radiation-induced spirals. To quantify this effect, we run a test where we take our \revision{fiducial} low-resolution simulation \revision{for a Saturn-mass planet} at 1000 orbits, institute an upper wave-damping boundary with an aggressive damping time $t_{\rm damp} = 5 \times 10^{-3}$ orbits, and run it for another 10 orbits---enough to see changes to the spiral arms.

\begin{figure}
    \centering
    \includegraphics[width=0.5\textwidth]{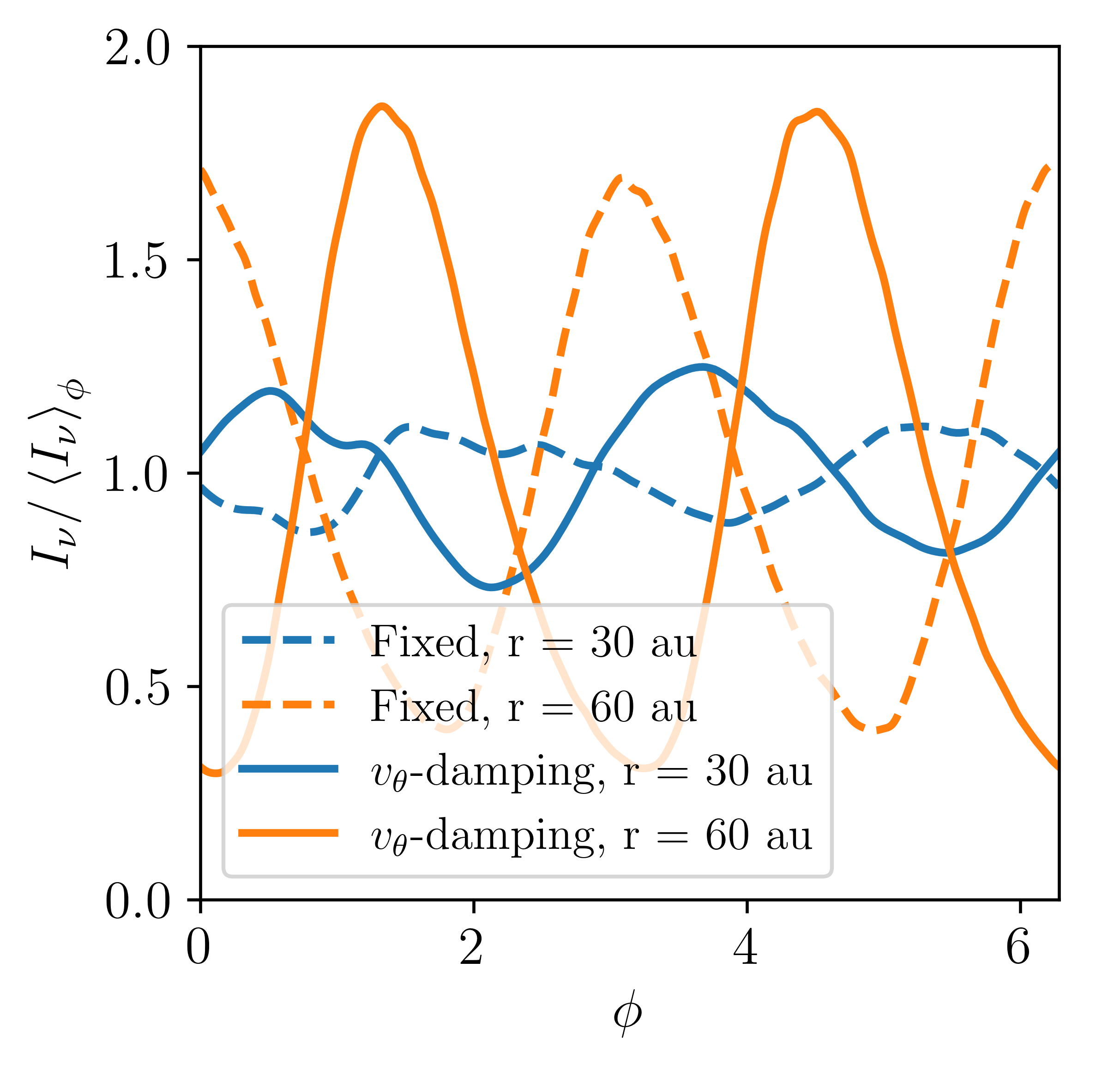}
    \caption{$H$-band contrast ratios from post-processed simulations with the standard fixed boundary condition, as well as the strong-damping condition.}
    \label{fig:contrast_bc}
\end{figure}

In Figures \ref{fig:h_band_bc} and \ref{fig:contrast_bc} we make plots analogous to Figures \ref{fig:NIR_mock_images} and \ref{fig:contrast}, one the simulation at 1000 orbits with fixed boundaries, and one after 10 more orbits with the wave-damping zone implemented. The spiral is qualitatively similar, but qualitatively somewhat stronger, when reflective boundaries are used. An analogous test with the polar boundaries extended to $\theta = \pm 0.6$ above and below the midplane, as well as another with the fiducial $\theta$-extent, but where reflective polar boundaries are used from the start---both not shown here---show similar results. This demonstrates the robustness of our conclusions with respect to numerical prescription, and clears the way to use reflecting boundary conditions to model upper-disk and midplane dynamics simultaneously.

%\end{comment}

\end{document}